\documentclass{emulateapj}

\usepackage{graphics,graphicx}
\usepackage{natbib}
\usepackage{epstopdf}
\def\apj{\rm ApJ}
\def\apjl{\rm ApJL}

\def\mnras{\rm MNRAS}
\def\nat{\rm Nature}

\newcommand{\gtorder}{\mathrel{\raise.3ex\hbox{$>$}\mkern-14mu
            \lower0.6ex\hbox{$\sim$}}}
\newcommand{\ltorder}{\mathrel{\raise.3ex\hbox{$<$}\mkern-14mu
            \lower0.6ex\hbox{$\sim$}}}

\shorttitle{SMBH formation in stellar clusters}
\shortauthors{Davies et al.}

\begin{document}

\title{SMBH formation via gas accretion in nuclear stellar clusters}

\author{Melvyn B. Davies\altaffilmark{1}, M. Coleman Miller\altaffilmark{2},
Jillian M. Bellovary\altaffilmark{3}}

\altaffiltext{1}{Lund Observatory, Box 43, SE--221 00, Lund, Sweden}
\altaffiltext{2}{Department of Astronomy and Joint Space-Science Institute, 
University of Maryland, College Park, MD}
\altaffiltext{3}{Department of Astronomy, University of Michigan, Ann Arbor, MI}

\begin{abstract}

Black holes exceeding a billion solar masses have been detected at
redshifts greater than six.  The rapid formation of these objects may
suggest a massive early seed or a period of growth faster than
Eddington.  Here we suggest a new mechanism along these lines.  We
propose that in the process of hierarchical structure assembly, dense
star clusters can be contracted on dynamical time scales due to the
nearly free-fall inflow of self-gravitating gas with a mass comparable
to or larger than that of the clusters.  This process increases the
velocity dispersion to the point where the few remaining hard binaries
can no longer effectively heat the cluster, and the cluster goes into
a period of homologous core collapse.  The cluster core can then reach
a central density high enough for fast mergers of stellar-mass black
holes and hence the rapid production of a black hole seed that could
be $10^5~M_\odot$ or larger.

\end{abstract}


\section{Introduction}

Evidence from observations of high redshift active galactic nuclei
(AGN) shows that massive black holes ($m_{\rm bh} \sim 10^9$
M$_\odot$) have already formed at red shifts $z>6$ \citep{Willott03,
Barth03}.  It is challenging to produce such masses so early via
Eddington-limited gas accretion onto initially stellar-mass black
holes, hence there have been various explorations of the possibility
of higher-mass seeds, e.g., Population III stars \citep{Madau01,
Volonteri03}, massive stars formed through run-away collisions
\citep{Devecchi09} or quasi-stars \citep{Begelman06, Lodato06}.  Here
we renew a earlier suggestion: that a
sufficiently massive and dense star cluster with a central black hole
sub-cluster could undergo collapse, including mergers due to
gravitational radiation, building up a massive black hole much
faster than would be possible by Eddington-limited gas accretion onto
a stellar-mass seed.

SMBH formation within stellar clusters in galactic nuclei has been
explored in a series of papers by \citet{Quinlan87, Quinlan89,
Quinlan90}. They considered core collapse in a cluster of compact
objects where the cluster core contracts driven by two-body relaxation
 to the point where the central potential well is so deep that a
relativistic instability sets in and a single black hole is formed via
the collapse of the central core of compact objects. However, this
instability requires a central redshift $z\sim 0.5$, meaning that the
core would have to be extremely massive and dense.

In this {\em Letter} we consider a variation on the cluster-collapse
model.  We begin with cluster conditions similar to those seen in
nuclear stellar clusters, which have masses and radii roughly
comparable to globular clusters (e.g., \citealt{Seth08}).  In these
clusters, the more-massive stellar-mass black holes are likely to
segregate to the cluster center forming their own dark core.  Binaries
within the core will provide the cluster with energy via binary-single
encounters.  These binaries are relatively wide ($\sim 0.1-1$ AU),
hence the timescale for them to merge via the effects of gravitational
radiation is extremely long. Binary heating will thus provide a fuel
to support the cluster core from complete collapse.

Infall of gas into such a nuclear stellar cluster may have profound
consequences for its evolution. Infall is likely to occur during
a merger between two galaxies \citep{Mayer10}, which will be common at
high redshifts \citep{Bellovary11}.  The addition of significant mass
will cause the cluster to shrink after virialization, and in some
circumstances significant accretion onto cluster stars may occur. The
cluster potential well will be deepened, increasing velocity
dispersions and decreasing the number and semi-major axes of the
binaries that heat the cluster. Because the inspiral timescale due to
gravitational radiation depends sensitively on the initial separation,
these smaller binaries will have significantly decreased merger
timescales.  Mergers remove binaries and thus remove the energy
source for the cluster.  Without an energy source, the cluster core
will undergo deep core collapse, potentially causing a runaway merger
of the black holes residing there.  For this collapse to occur we
require not only that the merger timescale of binaries due to the
emission of gravitational radiation is small, but also that the merger
products (which will receive kicks due to the asymmetric emission of
gravitational waves) are retained within the cluster core.

In Section 2 we set out the conditions for a nuclear stellar cluster
such that black-hole binaries within the core are likely to merge (and
be retained) on shorter timescales than cluster heating via
binary-single encounters. In Section 3 we consider the immediate
effects of infall gas into a nuclear stellar cluster. In Section 4 we
consider the subsequent evolution of a nuclear stellar cluster after
an episode of gas infall, possibly leading to a chain of events
producing an SMBH.

\section{Dry nuclear stellar clusters}

We begin by considering a cluster free of gas and assume
the cluster contains only two species: stellar-mass black
holes, and less-massive main-sequence stars.  The heavier black holes
will sink to the core by the effects of mass segregation. The cluster
will likely be vulnerable to the Spitzer instability, whereby the
black holes will form their own central sub-cluster within the cluster
core.  Although some black hole -- black hole
interactions can lead to ejections, a significant fraction of the
initial black hole population remains even after several Gyr at
standard cluster densities \citep{Mackey2007,Mackey2008}.

Black hole binaries within the dark core that are soft (ie
their binding energy is less than the typical kinetic energy of a
single black hole) will be split rapidly by binary-single
interactions.  In contrast, hard binaries tend to harden further after
binary-single interactions (and heavier black holes tend to swap into
the binaries), meaning that the single and binary both get a kick
after the interaction and the cluster is heated as a result
\citep{Heggie75}.  The timescale for a given object to have a
gravitationally focused encounter with another object within the
cluster is given by \citep{Binney08}

\begin{equation}
\tau_{\rm enc} = 7 \times 10^{10}  n_5^{-1} v_{\infty,10} r_{\rm min}^{-1}
m^{-1} \ {\rm yr}
\end{equation}
where $n_5$ is the number density of stars/black holes in units of
$10^5$ objects/pc$^3$, $r_{\rm min} $ is the minimum distance during
the encounter in solar radii, $m$ is the sum of the masses of the
objects involved in the encounter in solar masses, and $v_{\infty,10}$
is the relative speed at infinity in units of 10km/s.  When
considering encounters between binaries and single stars, it is
reasonable to set $r_{\rm min} = a_{\rm bin}$, the semi-major axis of
the binary.  If we consider encounters between binary and single black
holes, all of equal mass, then $m = 3 m_{\rm bh}$. The number density
of objects in the core is $n \simeq 3 N_{\rm bh} / 4 \pi r_{\rm c}^3$,
where $N_{\rm bh}$ is the total number of black holes and $r_{\rm c}$
is the core radius. Assuming the cluster is in virial equilibrium, one
may estimate that $v_\infty \simeq \sqrt{0.4 G M_{\rm c} /r_{\rm h}}$
\citep{Binney08}, where $M_{\rm c}$ is the total cluster mass. This
implies $v_{\infty,10} \simeq 4.36 \sqrt{M_{\rm c,6}/r_{\rm h}}$ where
$M_{\rm c,6}$ is the total cluster mass in units of 10$^6$ M$_\odot$
and $r_{\rm h}$ is the cluster half-mass radius in pc.

The semi-major axis of a binary is given by

\begin{equation}
a_{\rm bin} \simeq { 1000 {\rm R}_\odot  \over x } \left( m_{\rm bh} \over 
v_{\infty,10}^2 \right)
\end{equation}
where $x$ is the ratio of the binary binding energy to the kinetic
energy of the stars/black holes, and $m_{\rm bh}$ is given in solar
masses. One can show that the timescale for
an encounter between a given black-hole binary and a single black hole
is given by

\begin{equation}
\tau_{2+1} \simeq  5 \times 10^{14} {r_{\rm c}^3 \over N_{\rm bh}} 
{x \over m_{\rm bh}^2} \left( M_{\rm c,6} \over r_{\rm  h} \right)^{3/2}  \ {\rm yr} 
\end{equation}
where $r_{\rm c}$ is the cluster radius in pc.
Setting $r_{\rm c} = \beta r_{\rm h}$, 
$m_{\rm bh} = 10 {\rm M}_\odot$, $m_\star = 1 {\rm M}_\odot$, 
and  $N_{\rm bh} = \alpha N_\star$, where 
$N_\star$ is the total number of stars in the cluster, we obtain

\begin{equation}
\tau_{2+1} \simeq 10^9 {\beta^3 x \over \alpha} { M_{c,6}^2 \over v_{\infty,10}^3} \  {\rm yr}
\end{equation}
Assuming a Salpeter IMF, with minimum stellar mass of $0.2$
M$_\odot$ and that all stars more massive than $25$ M$_\odot$ produce
black holes, we have $\alpha=0.0015$. Setting $\beta =0.1$ gives

\begin{equation}
\tau_{2+1} \simeq 6 \times 10^8 x { M_{c,6}^2 \over v_{\infty,10}^3} \  {\rm yr}
\end{equation}

Tight binaries will lose energy via the emission of gravitational
radiation. The timescale for two black holes, in a binary of initial
separation $a_{\rm bin}$ and eccentricity $e$, to spiral together is
given by \citep{Peters64}

\begin{equation}
\tau_{\rm gr} \simeq 10^{10}  \left( a_{\rm bin} \over 3.3 {\rm R}_\odot \right)^4
\left(1 \over 2 m_{\rm bh}^3 \right) 
\left( 1 - e^2 \right)^{7/2}  \ {\rm yr}
\end{equation}

Binary-single scattering will leave the binaries with a thermal
distribution of eccentricities where the distribution of orbital
eccentricities is given by $dn/de =2e$. The median eccentricity is
thus $e_{\rm med} = 1/\sqrt{2}$, hence a typical binary merger time is
reduced by a factor $\sim 10$ and could be reduced by much more for
higher eccentricities.  Using the expression for $a_{\rm bin}$ given
above, we obtain

\begin{equation}
\tau_{\rm gr} \simeq {5\times 10^{19}} {m_{\rm bh} \over v_{\infty,10}^8 }
x^{-4} \left( 1 - e^2 \right)^{7/2}  \  {\rm yr}
\end{equation}

\begin{figure}[htb]
\begin{center}
\plotone{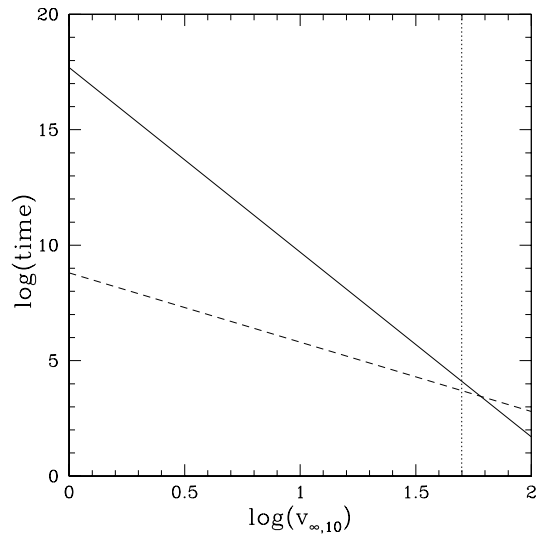}
\caption{Log of timescales (in years)  for binary-single encounters
(dashed line) and gravitational radiation inspiral (solid line) in
binaries as a function of log of velocity dispersion within clusters,
$v_{\infty,10}$, in units of 10km/s, assuming  binaries with 
hardness parameter $x=3$, and eccentricity $e=1/\sqrt{2}$. The vertical dotted
line  represents the value of $v_{\infty,10}$ when merging black
holes are likely to be retained.}
\label{fig:dmb1}
\end{center}
\end{figure}

We plot $\tau_{\rm gr}$ and $\tau_{2+1}$ as a function of
$v_{\infty,10}$ in Figure~\ref{fig:dmb1}, taking $x=3$, and assuming
the binaries initially have the median eccentricity of a thermal
distribution ($e=e_{\rm med}= 1/\sqrt{2}$).  
Thus, for moderately hard ($x \sim 3$) binaries in a typical nuclear stellar
cluster ($v_{\infty,10} \sim 3$), $t_{\rm gr}$ is extremely large:
black hole binaries are unlikely to merge via the effects of
gravitational radiation.  If $\tau_{\rm gr} < \tau_{2+1}$,
binaries will merge before they have the chance to heat the cluster.
The exact value of $v_{\infty,10}$ when this is the case will depend
on the hardness and eccentricity of the binary, but for moderately
hard, and eccentric, binaries, this is likely to be true for
$v_{\infty,10} \gtorder 50$.

When two inspiraling black holes merge, they receive a kick due
to the asymmetric emission of gravitational radiation. This kick
depends on the mass ratio of the two black holes and their spins, but
can be as large as $\sim 4000~{\rm km~s}^{-1}$ for an optimal
configuration (e.g., \citealt{Baker08}).  In order to retain merger
products within the stellar cluster, we require the cluster
escape speed to exceed 1000 km/s (or $v_{\infty,10} \sim 50$). This is
equivalent to requiring that

\begin{equation}
\left(M_{\rm c,6} \over r_{\rm h} \right)^{1/2}  \ge 10
\label{eqn:dmb8}
\end{equation}
Any merger products ejected {\it extremely close} to the escape
speed may be left on very wide orbits outside of the cluster, however
objects ejected at less than about 80 \% of the escape speed will
remain in the cluster.  We thus see that the conditions required for
the retention of merger products are rather similar to those which
give us $\tau_{\rm gr} < \tau_{2+1}$.  If the merger products are
retained within the cluster, we 
expect the dark core to continue to contract, having the potential to lead to a runaway merger
of black holes. The outcome of such a process could be the production
of an intermediate-mass black hole which would act as the seed for an
SMBH.

However, most observed nuclear stellar clusters will {\it not} satisfy
equation~(\ref{eqn:dmb8}).  We therefore require some mechanism to
cause the cluster to shrink and/or increase its mass in order to reach
the conditions required for runaway black hole mergers to occur within
the dark core.  

\section{The effect of gas infall into the cluster}

Suppose that gas with mass comparable to or larger than the
stellar mass of the cluster falls quickly to the center of the
cluster.  Just as in the inverse problem (e.g., where mass is expelled
from a binary in a supernova), we assume that the angular momentum of
individual stellar orbits is conserved, meaning that the
characteristic size of the orbits is proportional to $1/M$, where $M$
is the mass interior to the orbit.  Thus, for example, adding a gas
mass equal to the cluster mass halves the sizes of the orbits exterior
to the gas and increases the stellar density by a factor of eight.  As
shown below, at redshifts $z>10$ one can indeed have gas inflows
with total masses up to $\sim 10$ times the mass of the stellar
cluster.  The key question is whether much of the gas flows to the
inner $\sim$pc or closer, rather than stalling and forming stars much
farther away, which would not contract the central cluster.

One might imagine that a cooling flow could form that would bring the
gas in.  Indeed, such a scenario has been suggested by
\citet{Vesperini10} to grow intermediate-mass black holes in early
globular clusters.  However, as demonstrated by \citet{D'Ercole08},
the distributed injection of even a comparatively small amount of
luminosity ($Q_{\rm cr}\sim 6\times 10^{37}~{\rm erg~s}^{-1}$ in their
treatment) is sufficient to hold off the cooling flow.  This is
considerably less than the Eddington luminosity from even a single
$10~M_\odot$ black hole, and given that this luminosity is typically
distributed over a large spread in photon energies that is therefore
absorbed over a wide range in gas column depths, it appears unlikely
that a cooling flow will develop in the situation we consider here.

We instead turn to the scenario proposed by \citet{Mayer10}, in which
the self-gravitating gas is subject to instabilities that funnel much
of the gas to the center in a low angular momentum flow that, in their
calculations, gets to 0.2~pc or closer to the center.  
Gravitationally driven turbulence prevents the gas from fragmenting
into stars.  Such a flow may lead to the formation of a massive black
hole, as \citet{Mayer10} suggest, but the large range of scales
between the final $\sim 0.1$~pc simulated by \citealt{Mayer10} and the
$\sim 0.01$~AU scale of the black hole means that the final collapse
is still unresolved and might be hampered by a variety of effects.
Nonetheless, the inflow will also very effectively contract an
existing stellar cluster, leading to high central densities, rapid
mass segregation, and fast interactions among the existing
stellar-mass black holes that could lead to quick coalescence and the
formation of a massive black hole seed.  This scenario is particularly
probable given that such processes likely occurred in miniature in the
smaller clusters and gas flows that accumulated to form the central
cluster.  Thus the initial black hole masses would have had a wide
range; this is known to speed up mass segregation and coalescence
\citep{Quinlan87}.

\begin{figure}[htb]
\begin{center}
\plotone{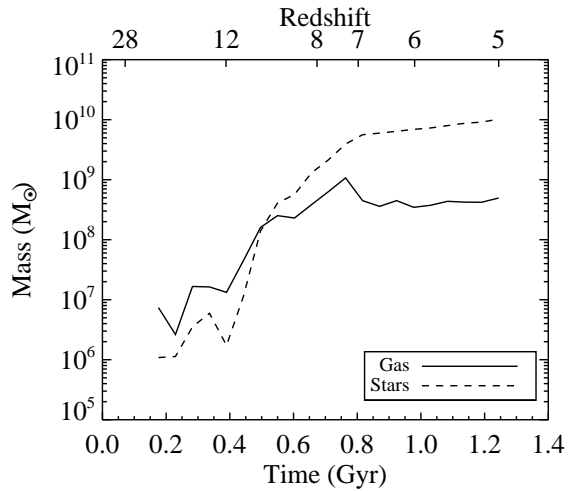}
\caption{Amount of mass as a function of time for gas (black
line) and stars (dashed line) in the cosmological simulation.  Masses
are the total mass enclosed within 520 pc (two softening lengths) from
the center of the primary galaxy in the simulation.}
\label{fig:dmb2}
\end{center}
\end{figure}

As an illustration of the gas infall into nuclear regions which can
occur, we turn to high-resolution cosmological simulations of galaxy
formation, considering the evolution of a massive galaxy destined to
become a massive elliptical at low redshift \citep{Bellovary11}.  In
Figure~\ref{fig:dmb2} we show the mass enclosed within 520 pc 
of the galaxy center vs time for the
gas (black line) and stars (dashed line).  The gaseous and stellar
inflow are due to a combination of direct accretion of matter from the
cosmic web/filaments as well as galaxy mergers. From this plot we see
that significant inflow of both gas and stars occurs at high redshift.
Repeating this measurement for similar simulations reveals that high
redshift ($z>10$) massive inflow on this scale is a common
occurrence for galaxies in this mass range (a few $\times 10^{11}$
M$_\odot$ at $z = 5$).

One might wonder whether the rapid contraction of a cluster by a
factor of a few would lead to a high rate of interactions of single objects with hard
binaries and hence to re-expansion of the cluster due to energy input.
The degree to which this can happen clearly depends on the fraction of
stars or black holes in hard binaries that are not so hard that they
merge or collide rapidly.  This fraction, in turn, depends on the
velocity dispersion; higher velocity dispersion means fewer hard
binaries and less binary binding energy that can potentially be tapped
to hold off core collapse.  We also note that close three-body
interactions between objects of comparable mass yield a thermal
distribution of eccentricities.  The eccentricities that give
pericenter distances low enough for collisions ($\sim 0.01$~AU for
solar-type stars) or fast merger by gravitational radiation (also
$\sim 0.01$~AU for $\sim 10~M_\odot$ black holes to merge in a few
million years or less) will destroy the binary.  
If the hard-soft boundary is $a<1$~AU
(corresponding to a velocity dispersion $\sigma\sim 30$~km~s$^{-1}$
for solar-type stars or $\sigma\sim 100$~km~s$^{-1}$ for $\sim
10~M_\odot$ black holes), the available binding energy in binaries is
less than the binding energy in the singles, hence binary-single
interactions are inefficient at holding off core collapse.

\section{The fate of the mass-loaded cluster}

We now consider the subsequent evolution of the stellar cluster,
assuming it has received a significant influx of gas, and perhaps
stars. Beginning with a nuclear stellar cluster of mass $M_{\rm
c,6}æ\sim 1$, and half-mass radius $r_{\rm h} \sim 1$, the infall of
$10^7$ M$_{\odot}$ of gas is likely to shrink the cluster by a factor
of about ten leaving initial core densities $n \sim 10^8$ stars
pc$^{-3}$ and velocity dispersions $v_{\infty,10} \sim 30 -
50$. Several things will happen to this cluster.  Wider binaries
(which previously heated the cluster and prevented core collapse) will
now be soft, as the velocity dispersion has increased, and will be
broken up. Any remaining binaries will quickly merge via gravitational
radiation inspiral or collisions. Merger products are likely to be
retained with the cluster owing to the increased escape speed which
resulted from the gas infall.  It is possible that some merger
products will in turn exchange into other binaries which then
merge. This process may even be repeated.

After a short time, the cluster core will contain black holes having a
broader range of masses than previously. In addition, all binaries
will either have been broken up or have merged.  Thus the cluster core
will lose its source of energy, and will begin to contract approaching
core collapse as it transfers energy to the cluster halo via two-body
scattering.  As the core density increases, runaway mergers between
any non-compact stars within the cluster core becomes possible
\citep{Quinlan90}. Such collisions may produce more massive stars which
in turn will add to the population of black holes and neutron stars.

At high densities, extremely close encounters occur between two
compact objects where binary formation via gravitational wave emission
is possible. The timescale for capture is given by \citep{Quinlan89}

\begin{equation}
\tau_{\rm cap} \approx 7\times 10^{12} 
n_5^{-1} \mu_{10}^{-2/7} M_{100}^{-12/7} v_{\infty,500}^{-11/7}  \ {\rm yr}\; ,
\end{equation}
where $\mu_{10}$ is the reduced mass of the two compact objects
capturing each other (in units of 10 M$_\odot$ and $M_{100}$ is their
total mass (in units of 100 M$_\odot$).  We see that when the core
reaches densities $n \sim 10^{12}$ stars pc$^{-3}$, which at a
velocity dispersion $v\sim 300~{\rm km~s}^{-1}$ contains ${\rm
few}\times 10^4~M_\odot$, the formation of BH-BH binaries via
gravitational wave emission becomes possible on a timescale
$\ltorder$Myr.  These binaries will have very short inspiral
timescales.  We will thus have a second phase of black hole binary
mergers.  The black holes produced via such mergers have greater mass,
and given that the timescale for mergers scales as $M^{-12/7}$,
higher-mass black holes have a greater probability of merger and the
process will run away. Such a phase may well collect most of the mass
of the stellar black holes into a single black hole, which would have
several percent of the original stellar mass of the system for typical
initial mass functions.  Thus the mass of the single black hole could
be $\sim 10^5$ M$_\odot$ for situations similar to that shown in
Figure~\ref{fig:dmb2} (assuming most infalling gas reaches the centre).
  Accretion of neutron stars and white dwarfs,
both of which will be swallowed whole for $M\gtorder 10^5~M_\odot$,
could boost the mass by a factor of a few, depending on the mass at
which stellar interactions in the radius of influence of the black
hole become an effective heating source for the cluster (e.g.,
\citealt{Gill08}).  Overall, 
the time elapsed between
mass infall to the production of the seed SMBH will be roughly 100~Myr
(e.g., \citealt{Quinlan90}), hence producing a $\sim 10^5~M_\odot$
black hole within $\sim 300-400$~Myr after the beginning of the
universe.  Growth to $\sim 10^9~M_\odot$ via gas accretion could then
occur by a redshift $z\sim 6$ via standard Eddington-limited accretion
onto moderately spinning black holes.

In summary, we have proposed that clusters with initial structural
parameters similar to current-day nuclear clusters may (1)~form as
part of early hierarchical merging, (2)~accrete gas with a total mass
comparable to or greater than that of the cluster at redshifts $z>10$,
(3)~contract as a result so that binary heating is ineffective,
(4)~undergo core collapse to a density high enough that stellar-mass
black holes merge, and thus (5)~have most of the mass originally in
stellar-mass black holes collect in to a single black hole that could
be $\sim 10^5~M_\odot$ or larger.  This black hole would therefore be
a high-mass seed that could comfortably grow to supermassive size by
the observed redshifts $z\sim 6$.

\acknowledgements

We gratefully acknowledge the hospitality of the Aspen Center for
Physics.  MBD was supported by 
the Swedish Research Council (grant 2008-4089).
 MCM was supported in part by NASA grant NNX08AH29G.  JMB
acknowledges NASA award NNX10AC84G.

\end{document}